\documentclass[aps,prb,twocolumn,showpacs,preprintnumbers,amsmath,amssymb,superscriptaddress, 10pt]{revtex4-1}%
\usepackage{graphicx}
\usepackage{dcolumn}
\usepackage{amsmath}
\usepackage{color}
\usepackage{bm}% bold math

\newcommand{\tnfe}{\textit{T}$_{\rm N}^{\rm Fe}$}
\newcommand{\tnull}{\textit{T}$_0$}

\begin{document}
\title{Anisotropic electrical resistivity of LaFeAsO: \\evidence for electronic nematicity}

\author{A. Jesche}
 \email[]{jesche@ameslab.gov}
 \affiliation{The Ames Laboratory, Iowa State University, Ames, USA}
 \affiliation{Max Planck Institute for Chemical Physics of Solids, D-01187 Dresden, Germany}
\author{F. Nitsche}
 \affiliation{Department of Chemistry and Food Chemistry, Technische Universit\"at Dresden, D-01062 Dresden, Germany}
\author{S. Probst}
 \altaffiliation{Present address: Physikalisches Institut, Karlsruhe Institute of Technology, D-76128 Karlsruhe, Germany}
 \affiliation{Department of Physics, Universit\"at Erlangen-N\"urnberg, D-91058 Erlangen, Germany}
\author{Th. Doert}
 \affiliation{Department of Chemistry and Food Chemistry, Technische Universit\"at Dresden, D-01062 Dresden, Germany}
\author{P. M\"uller}
 \affiliation{Department of Physics, Universit\"at Erlangen-N\"urnberg, D-91058 Erlangen, Germany}
\author{M. Ruck}
 \affiliation{Department of Chemistry and Food Chemistry, Technische Universit\"at Dresden, D-01062 Dresden, Germany}
 \affiliation{Max Planck Institute for Chemical Physics of Solids, D-01187 Dresden, Germany}

\begin{abstract}
Single crystals of LaFeAsO were successfully grown out of KI flux. Temperature dependent electrical resistivity was measured with current flow along the basal plane, $\rho_\perp(T)$, as well as with current flow along the crystallographic $c$-axis, $\rho_\parallel(T)$, the latter one utilizing electron beam lithography and argon ion beam milling. The anisotropy ratio was found to lie between $\rho_\parallel$/$\rho_\perp = 20 - 200$. 
The measurement of $\rho_\perp(T)$ was performed with current flow along the tetragonal [1\,0\,0] direction and along the [1\,1\,0] direction and revealed a clear in-plane anisotropy already at $T \leq 175$\,K. This is significantly above the orthorhombic distortion at \tnull~=147\,K and indicates the formation of an electron nematic phase.
Magnetic susceptibility and electrical resistivity give evidence for a change of the magnetic structure of the iron atoms from antiferromagnetic to ferromagnetic arrangement along the $c$-axis at $T^* = 11$\,K.  
\end{abstract}

\maketitle
\section{Introduction}
The discovery of superconductivity in fluorine-doped LaFeAsO by Kamihara\,\textit{et al.}\,\cite{Kamihara2008} led to the finding of several other iron-based superconductors.
A common feature among these materials is an antiferromagnetic (AFM) ordering of iron in parent compounds that has to be sufficiently suppressed to induce superconductivity\,\cite{Johnston2010, Stewart2011}. 
Furthermore, there is increasing evidence for the relevance of an electron nematic phase on the emergence of superconductivity\,\cite{Fang2008}.
Thereby, the electronic symmetry is broken when compared to the underlying crystallographic symmetry, i.e., the 4-fold rotation symmetry along the crystallographic $c$-axis is broken in the tetragonal phase of the Fe-based superconductors. 
Electronic nematicity has been intensely studied in $A$Fe$_2$As$_2$ compounds ($A$ = Ca, Ba, Sr), mainly by measuring the in-plane anisotropy of the electrical resistivity\,\cite{Tanatar2010, Chu2010, Blomberg2011, Fisher2011, Blomberg2012}.
Recent theoretical work by Fernandes\,\textit{et al.} showed that the observed in-plane anisotropy can be well described within a nematic scenario\,\cite{Fernandes2011}.  
Further evidence for a broken in-plane symmetry stems from inelastic neutron scattering\,\cite{Lester2010}, elastic properties\,\cite{Fernandez2010}, optical spectroscopy\,\cite{Nakajima2011, Dusza2012}, angle resolved photo emission spectroscopy\,\cite{Wang2010-arxiv, Yi2011}, and magnetic torque measurements\,\cite{Kasahara2012}.
These results could be obtained because good quality single crystals are available for the $A$Fe$_2$As$_2$ compounds. 
In contrast, the synthesis and in particular the single crystal growth for the other large family of Fe-based superconductors, the $R$FeAsO compounds with $R$ = rare earth metal, is still more challenging\,\cite{Karpinski2009, Yan2011}.
Accordingly, the situation is much less clear and the relevance of nematic fluctuations for the $R$FeAsO is not settled.
Indications for a breaking of the 4-fold rotation symmetry in the tetragonal phase of LaFeAsO can be inferred from structural and elastic properties which show a 'gradual orthorhombic distortion'\,\cite{Mcguire2008, Qureshi2010} and a softening of elastic moduli\,\cite{Mcguire2008}, respectively, for cooling below $T = 200$\,K which is well above the proposed structural transition temperature of \tnull $~\approx 160$\,K. 

Here we present results on the anisotropic electrical resistivity of LaFeAsO giving direct evidence for a broken in-plane symmetry and show that nematic fluctuations are a general property of both $A$Fe$_2$As$_2$- and $R$FeAsO-type compounds.

Furthermore, we discuss the sample dependence of the electrical resistivity of LaFeAsO which seems to be particularly strong when compared to other members of the $R$FeAsO family, see e.g., Ref.\,\onlinecite{Kamihara2008, Mcguire2008, Shekhar2009, Yan2009-arxiv}. 

\section{Experimental}

\begin{figure}[!]
\includegraphics[width=0.4\textwidth]{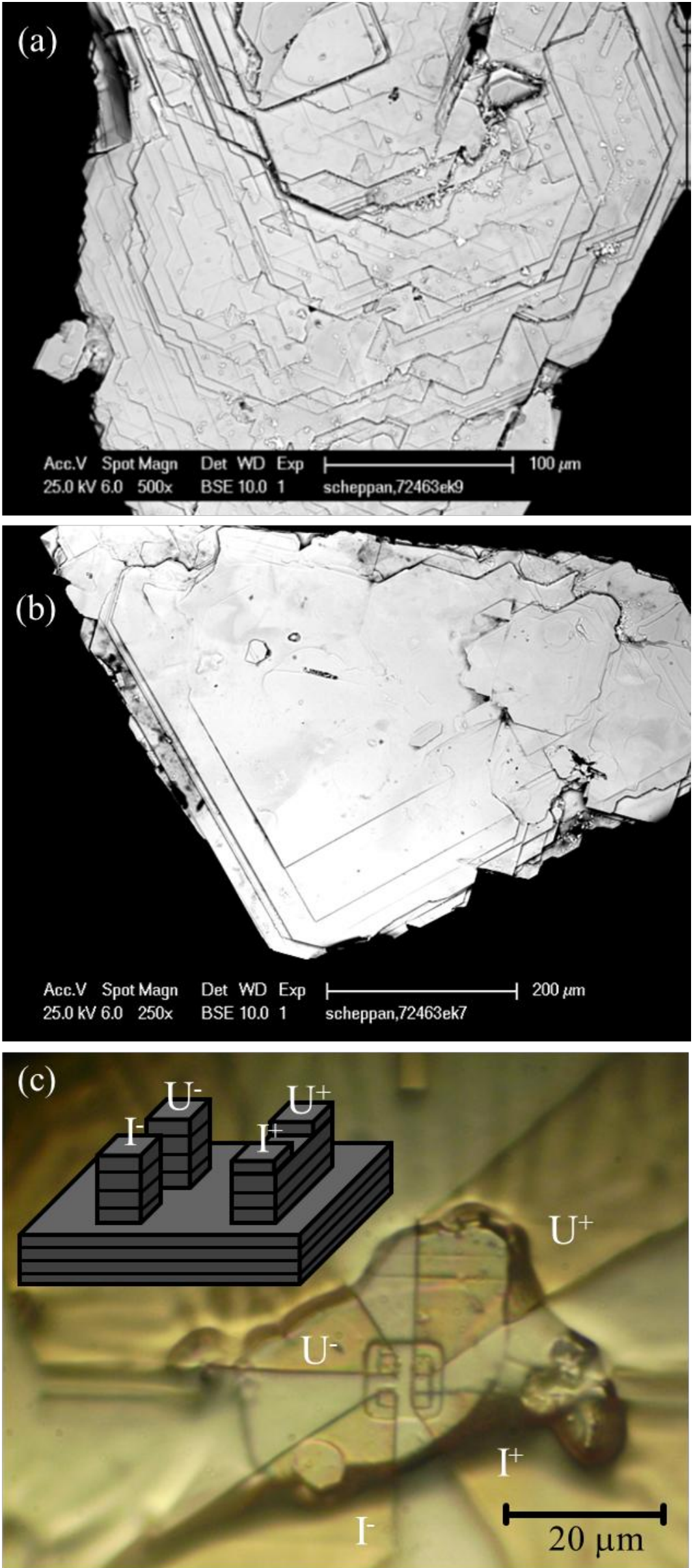}
\caption{(color online) (a,b) Electron micrographs of single crystalline LaFeAsO: (a) pronounced terraces in sample 1, (b) flat surface of sample 4. (c) A LaFeAsO crystal (sample 5) with two $2 \times 2\,\mu m^2$ sized tower-like structures (mesas) on the left and one rectangular $2 \times 6\,\mu m^2$ mesa with two $2 \times 2\,\mu m^2$ mesas on top on the right hand side. 
This micrograph shows the sample before its final argon ion milling step. The sample is covered with silver and the future positions of the contact leads are covered with photoresist.
The layout enables a four-point $c$-axis transport measurement on the $2 \times 6\,\mu m^2$ mesa. The height of this mesa is approximately 240 nm. A schematic drawing is shown in the upper left.}
\label{rem}
\end{figure}

Large LaFeAsO single crystals were grown by a high-temperature adaptation of the growth
conditions described in Ref.~\citenum{Nitsche2010}. The reactants are air-sensitive and were handled in an argon-filled glovebox (M. Braun,
$p(\text{O}_2)/p_0 \leq 1\,$ppm, 
$p(\text{H}_2\text{O})/p_0 \leq 1\,$ppm, 
argon purified with molecular sieve and copper catalyst). Iron(II)-oxide ($99.9\,$\%, Sigma-Aldrich), arsenic ($99.999\,$\%, Alfa Aeser), and freshly
filed lanthanum ($99.9\,$\%, Treibacher) were mixed and transferred to a
glassy carbon crucible (diameter: $25\,$mm, height: $37\,$mm, wall thickness:
$3\,$mm). $300\,$wt.-\%\ of potassium iodide ($99.5\,$\%, Gr\"{u}ssing, dried
at $650\,$K in dynamic vacuum) was used as the flux and stacked below
and above the reactants. Subsequently, the crucible was enclosed in a tantalum
container and welded closed at $500\,$mbar argon atmosphere. 

The mixtures were heated to
$1570\,$K over $13$ hours in a
furnace with static argon atmosphere. After 5 hours, the furnace was slowly cooled down to $1370$\,K (batch 1) or $1070\,$K (batch 2) at $5\,$K$\,$/hour and finally cooled down to room temperature (RT) at $\sim200\,$K/$\,$hour (no systematic differences between batch 1 and batch 2 were observed).

The flux was removed by deionized water to isolate plate-shaped single crystals of up to a millimeter along a side. 
Electron micrographs of two LaFeAsO single crystals are shown in Fig.\,\ref{rem}(a,b). The surfaces can form pronounced terraces (a) or be rather flat (b).  
Note, the size of the single crystals is sensitive to the grain size of iron(II)-oxide; grinding the starting material to a fine powder resulted in significantly smaller LaFeAsO single crystals whereas the presence of iron(II)-oxide lumps ($\approx\,0.1 - 1$\,mm) favors the growth of larger samples. 
Further attempts to evaluate the influence of the solubility of FeO in KI-flux have been performed by cold-pressing ground FeO powder into a pellet and using this as starting material. After carrying out the growth procedure as described above, the original pellet form was still recognizable. However, this hard and stable pellet was found to consist of Fe$_2$As instead of FeO. The other products were La$_2$O$_3$, LaAs, and a minuscule amount of LaFeAsO. 
Using polycrystalline LaFeAsO and KI-flux as starting materials did not yield single crystals. These observations point to a non-equilibrium crystal growth with the details being thus far unclear.

Energy dispersive x-ray analysis revealed a stoichiometric La:Fe:As content and confirmed the presence of oxygen. The characteristic emission lines of potassium or iodine were not observed,
thus substantial inclusion or incorporation of the flux
material can be excluded.
Powder x-ray diffraction patterns were measured at $293(1)\,$K on a Stadi P diffractometer
(Stoe \&\ Cie., Cu~$K\alpha_1$, Ge monochromator). Lattice parameters were determined by LeBail pattern decomposition
%  with internal LaB$_6$ standard(NIST~660b) 
using GSAS\cite{Larson2000} and EXPGUI\cite{Toby2001}.
To orient the large single crystals and to assess their mosaicity, 
images of sections of the reciprocal space were recorded using a
Buerger precession camera (Huber, Mo anode, Zr filter).
Low temperature $\omega$ scans (1$^\circ$) of single crystals, cooled with an Oxford Cryostream 700, were collected with a Bruker SMART diffractometer (Mo~$K\alpha$, graphite monochromator).

Electrical resistivity for current flow along the basal plane, $\rho_\perp$, was measured in a 4-point geometry using the AC transport option of a Quantum Design Physical Property Measurement System (PPMS). Silver paint was used to connect Pt-wire to the as-grown plate-like samples [Fig.\,\ref{inplane}(b)].

To measure the electrical resistivity along the crystallographic $c$-direction, $\rho_\parallel$, mesa structures were created using electron beam lithography and argon ion beam milling on LaFeAsO single crystals [Fig.\,\ref{rem}(c)]. 
First, two quadratic and one rectangular mesa were etched out of a single crystal. Second, two additional mesas were designed on top of the rectangular mesa. In the last step, silver leads were structured on top in order to measure the $c$-axis transport of the rectangular mesa (for more detailed information see Ref.\,\onlinecite{Probst2011}). 
Note that this geometry has a small $ab$-plane contribution to the electrical resistance. However, its contribution is negligible as long as $\rho_\parallel >> \rho_\perp$.
The effective dimensions of the mesas were 2\,$\mu$m $\times$ 6\,$\mu$m $\times$ 0.24\,$\mu$m (sample 5) and 6\,$\mu$m $\times$ 18\,$\mu$m $\times$ 0.29\,$\mu$m (sample 6).
The $c$-axis transport measurements from $T = 284 - 4.2$\,K were performed using a self made dipstick setup inside a standard liquid He transport dewar. 

Magnetization measurements were performed using a Quantum Design Magnetic Property Measurement System (MPMS).
 
\section{Structural characterization}

\begin{figure}
\includegraphics[width=0.47\textwidth]{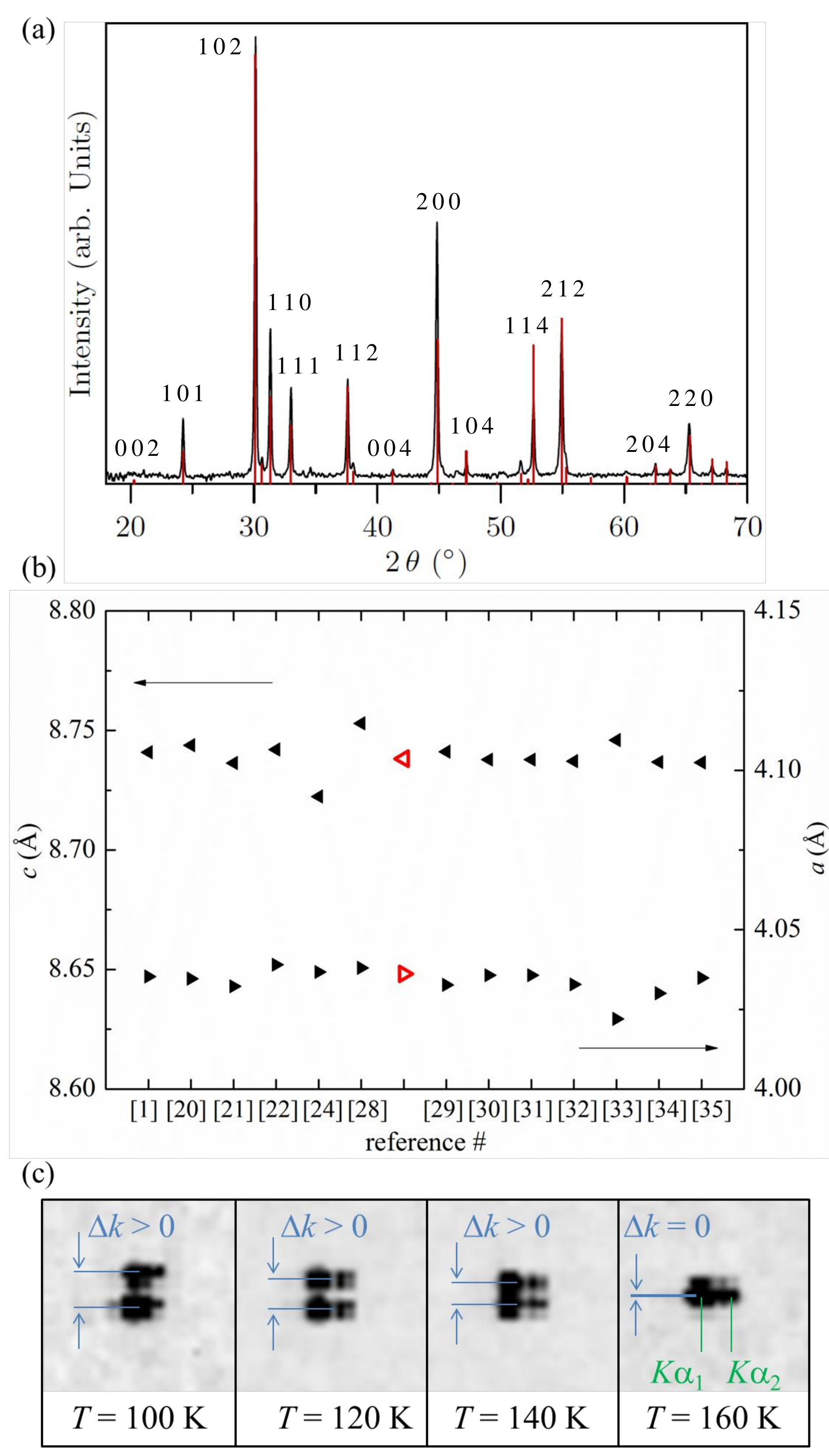}
\caption{(color online) (a)  X-ray powder diffraction pattern of ground LaFeAsO crystals. Red lines indicate calculated reflections for LaFeAsO.
(b) Lattice parameters obtained for LaFeAsO at RT (red, open triangles) in comparison with literature data (black triangles). (c) Splitting of the 800 Bragg reflection manifesting the orthorhombic distortion below \tnull~= 147\,K (the splitting in reciprocal space is labeled by $\Delta k$).}
\label{diffr}
\end{figure}

X-ray diffraction of large single crystals (Fig.\,\ref{rem}a,b and Fig.\,\ref{inplane}a,b)
revealed $\langle 1\, 1\, 0\rangle$ as the dominant growth directions for LaFeAsO.
The plate-shaped crystals exhibit mostly $\{1\,0\,0\}$, $\{0\,0\,1\}$ faces
with a high aspect ratio (typically above 20). The growth along
$\langle 0\, 0\, 1\rangle$ directions is much slower than along the
perpendicular directions. This leads to
unstable growth conditions perpendicular to the basal plane. Consequently, the
surface appears rough and a higher mosaicity can be observed for $0\,0\,l$
reflections. 
Reflections of reciprocal planes perpendicular to $[0\,0\,1]$ (e.g.
$h\,k\,1$ reflections in Fig.\,\ref{inplane}) are of low mosaicity and
show no significant misorientation parallel to the basal plane. Therefore, oriented
transport measurements are feasible on these large single crystals.

An x-ray powder diffraction pattern measured on ground single crystals is shown in Fig.\,\ref{diffr}a. The obtained lattice parameters of $a = 4.0361(4)$\,\AA~and $c = 8.7382(14)$\,\AA~are plotted in Fig.\,\ref{diffr}b in direct comparison with selected literature data 
(published standard deviations are smaller than the size of the symbols, the reference number is given on the abscissa: 
1 - Ref.\,\onlinecite{Kamihara2008},
2 - Ref.\,\onlinecite{Mcguire2008},
3 - Ref.\,\onlinecite{Qureshi2010},
4 - Ref.\,\onlinecite{Shekhar2009},
5 - Ref.\,\onlinecite{Nitsche2010},
6 - Ref.\,\onlinecite{Quebe2000},
7 - this paper,
8 - Ref.\,\onlinecite{Nomura2008},
9 - Ref.\,\onlinecite{Wang2009},
10 - Ref.\,\onlinecite{Guanghan2009},
11 - Ref.\,\onlinecite{Yanpeng2009},
12 - Ref.\,\onlinecite{Yan2009},
13 - Ref.\,\onlinecite{delaCruz2008},
14 - Ref.\,\onlinecite{Luo2009}). 
There is no significant deviation of our data from the statistical average of the lattice parameter. 
We emphasize this point to stress that the unprecedented change of the Fe magnetic structure (see below) is not a result of different lattice parameters.

However,
the influence of the As $z$-parameter on physical properties has not been studied systematically and can be responsible for the pronounced sample dependencies found for LaFeAsO in this as well as in other studies\,\cite{Yan2009-arxiv}. In fact, band structure calculations revealed a significant influence of the As $z$-parameter on electronic structure and magnetic properties\,\cite{Krellner2008b}. 
An accurate experimental determination of the As $z$-parameter can be done by means of single crystal x-ray diffraction. 
However, the single crystals used for electrical resistivity measurements are too large for this technique. An isolation of smaller suitable samples failed since the plate-like crystals are malleable and mechanical stress can easily lead to bending which causes broad reflection profiles.
Detailed results on the As $z$-parameters of $R$FeAsO ($R$ = La, Ce, Pr, Nd, Sm, Gd, and Tb) obtained on smaller single crystals were presented in an earlier publication\,\cite{Nitsche2010}.

Figure 2(c) shows the 8\,0\,0 reflection in the tetragonal phase at $T = 160$\,K (right hand side) and its splitting at $T = 140$\,K manifesting the structural distortion from tetragonal to orthorhombic symmetry in agreement with an ordering temperature of \tnull~=147\,K inferred from resistivity measurements (see below). 
The peak-broadening along the horizontal direction is caused by the presence of both Mo $K\alpha_1$ and Mo $K\alpha_2$ radiation.
The orthorhombic splitting further increases under cooling to $T = 120$ and 100\,K.
However, the temperature calibration of the setup ($\Delta T = \pm 5$\,K) is not sufficient for a quantitative analysis of the temperature dependence of the order parameter.

\section{Electrical resistivity}

\subsection{Anisotropy: \textit{ab}-plane vs. \textit{c}-axis}

\begin{figure}
\includegraphics[width=0.48\textwidth]{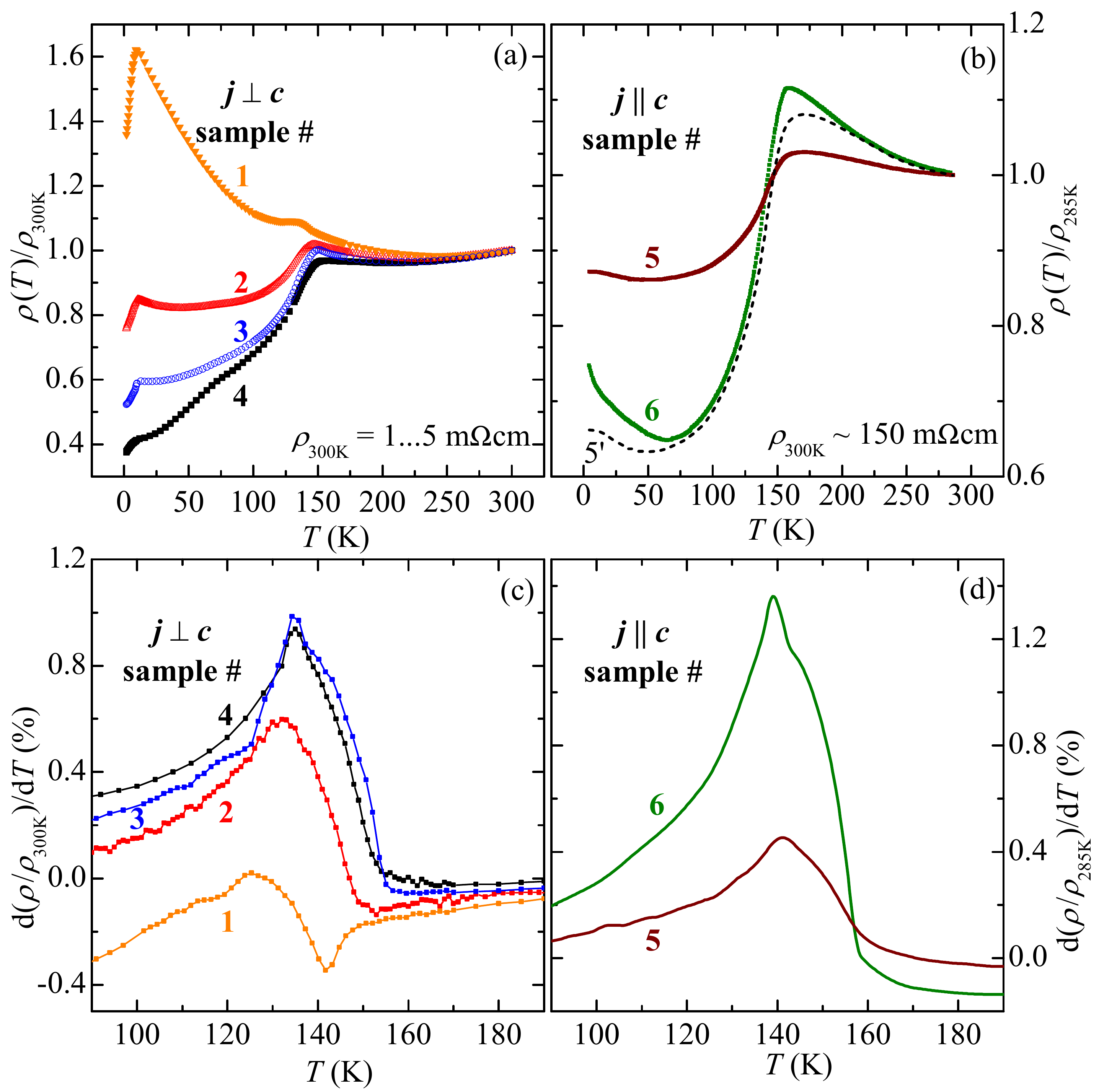}
\caption{
(Color online) Electrical resistivity of six LaFeAsO single crystals normalized to RT, (a) measured with current flow perpendicular to the $c$-axis for samples 1 - 4 and (b) parallel to the $c$ axis for samples 5 and 6 (5' - resistivity of sample 5 normalized to match the high-$T$ slope d$\rho_\parallel$/d$T\vert_{285 \rm K}$ of sample 6). 
A change of slope at \tnfe,\,\tnull $~\approx 135$\,K is observed in all samples for both orientations despite the significant sample dependencies at lower temperatures. A hitherto unobserved drop in $\rho(T)$ at $T = 11$\,K for current perpendicular to the $c$-axis is attributed to a change of the magnetic structure of Fe from AFM to FM arrangement along the $c$-axis. (c,d) Derivative of the normalized electrical resistivities manifest a maximum at \tnfe~and a shoulder towards higher temperatures associated with the structural distortion roughly 12\,K above \tnfe.
}
\label{rho}
\end{figure}

Fig.\,\ref{rho}a shows the electrical resistivity of four LaFeAsO single crystals from two different batches, all normalized to RT, with current flow perpendicular to the $c$-axis, $\rho_\perp$. Sample 2 and 4 belong to batch I, sample 1 and 3 belong to batch II.
The resistivity at RT varies from sample to sample between 1 to 5\,m$\Omega$cm without any correlation to the temperature dependence. Since the geometry factor is not well defined for these plate-like samples the absolute values are approximate.

$\bm{T > 150}${\,\bf K:}
All four samples exhibit a minimum in $\rho_\perp(T)$ shifting progressively from $T = 250$\,K for sample 1, to $T = 230$\,K for sample 2, to $T = 220$\,K for sample 3, and to $T = 205$\,K for sample 4.
The correlation with the low-$T$ behavior is evident: if the minimum occurs at higher temperatures and accordingly $\rho_\perp(T)$ increases strongly while cooling towards \tnull~then $\rho_\perp(T)$ shows a similar tendency to increase with decreasing $T$ at low temperatures. 

{\bf T $\approx$ 150\,K (\tnfe, \tnull):}
A sharp change of slope in $\rho_\perp(T)$ at $T = 150$\,K marks the structural transition and the AFM ordering of Fe in all samples.
The derivatives with respect to temperature, d$\rho_\perp$/d$T$, are plotted in Fig.\,\ref{rho}c.
The N\'eel temperature of Fe, determined by the maximum of the peak in d$\rho_\perp$/d$T$, was found to be \tnfe~= 135\,K for samples 3 and 4, which show relatively sharp maxima. For samples 1 and 2 the peak in d$\rho_\perp$/d$T$ is shifted to lower temperatures and is less pronounced but not significantly broadened. 
Taking into account the different $T$-dependence of $\rho_\perp(T)$ above and below \tnfe, it is presumably not a lower \tnfe~that causes the maximum in d$\rho_\perp$/d$T$ to occur at lower $T$ but the different 'background' of $\rho_\perp(T)$ in samples 1 and 2.
The peaks in d$\rho_\perp$/d$T$ for samples 3 and 4 have a shoulder towards higher temperatures that can be used to define the ordering temperature of the structural transition, \tnull~(see Ref.\,\onlinecite{Jesche2010}). 
For samples 3 and 4 we find \tnull = 147\,K which is $\sim12$\,K above \tnfe. Even though the shoulder in d$\rho_\perp$/d$T$ is less pronounced for samples 1 and 2, a similar splitting between \tnfe~and \tnull~can be inferred from the total width of the peak.

$\bm{T < 150}${\,\bf K:}
Under cooling the resistivity is either increasing (sample 1), decreasing (samples 3 and 4), or first decreasing followed by increasing (sample 2). 
Three possible explanations for this behavior are discussed in sec.\,\ref{DiscussionA}. 

A sharp drop in $\rho_\perp(T)$ at $T = 11$\,K has been observed in all samples. This behavior is reminiscent of NdFeAsO\,\cite{Tian2010}, where a similar anomaly in $\rho_\perp$ has been observed at $T^* = 15$\,K which is, as shown by neutron diffraction, associated with a change of the magnetic structure of Fe from AFM to ferromagnetic (FM) arrangement along the $c$-axis. 
For both AFM and FM arrangement along the $c$-axis, the Fe-moments are (anti)-parallel aligned along the orthorhombic $a$-axis with an AFM arrangement along the $a$-axis and a FM arrangement along the $b$-axis. In case of AFM arrangement along the $c$-axis a stripelike order forms and the magnetic unit cell is doubled along the $c$-axis when compared to the crystallographic unit cell. For FM arrangement along the $c$-axis the magnetic unit cell corresponds to the orthorhombic crystallographic unit cell.

Since the magnetic exchange along the $c$-axis seems to be frustrated in the $R$FeAsO systems, i.e., on the verge between AFM and FM, since the effective coupling changes from FM ($R$ = Ce\,\cite{Zhao2008}, Pr\,\cite{Kimber2008}, and Nd\,\cite{Tian2010} for $T < T^*$) to AFM ($R$ = La\,\cite{delaCruz2008}, Nd\,\cite{Tian2010} for $T > T^*$), we propose a similar change in the AFM structure of Fe in LaFeAsO at $T^* = 11$\,K. 
As will be discussed below, the measured magnetic susceptibility supports this interpretation. 
Note, we also synthesized CeFeAsO as described in the experimental section above and did not observe additional anomalies in $\rho(T)$ besides the ones at \tnfe~and $T_{\rm N}^{\rm Ce}$. Therefore, the use of KI-flux is not sufficient for the emergence of a $T^*$-anomaly which gives further evidence for the proposed scenario since the arrangement of the Fe-moments in CeFeAsO is already FM along the $c$-axis. 

Scanning electron microscope images revealed a rough surface for sample 1 with $ab$-planes forming distinct plateaus separated by sharp steps whereas sample 4 shows a rather smooth surface [Fig.\,\ref{rem}(a,b)]. Furthermore, a clear trend of a subsequently increasing roughness is observable from sample 1 to sample 4 (not shown). 
With the electrical contacts sitting on different $ab$-planes, a voltage drop along the $c$-axis can affect the measurement of $\rho_\perp$.
Therefore, we suspected a varying, semi-conducting $c$-axis contribution to the electrical resistivity to be the origin for the pronounced sample dependence. To prove this assumption the resistivity with current flow along the $c$-axis, $\rho_\parallel$, was measured for 2 samples from batch I. 
The resistivity at $T = 285$\,K was found to be $\rho_\parallel = 174(41)$\,m$\Omega$cm and $\rho_\parallel = 121(18)$\,m$\Omega$cm for sample 5 and sample 6, respectively, which is two orders of magnitude larger than $\rho_\perp$.
The temperature dependent resistivity normalized to $T = 285$\,K is plotted in Fig.\,\ref{rho}b. 
In contrast to $\rho_\perp(T)$ no minimum was observed in $\rho_\parallel(T)$ at $T \sim 250$\,K instead $\rho_\parallel(T)$ increases monotonously towards approaching \tnull.
The increase in $\rho_\parallel(T)$ of sample 6 at low-$T$ ($T < 70$\,K) is more pronounced than in sample 5 reflecting the high-$T$ behavior ($T = 150 - 285$\,K) corroborating the trend observed in $\rho_\perp(T)$.
However, if the resistivity of sample 5 is normalized to match the high-$T$ slope d$\rho_\parallel$/d$T\vert_{285 K}$ of sample 6, which can compensate for different residual resistivities and uncertainties in the geometry factor, the temperature dependence of both samples becomes similar, as seen in Fig.\,\ref{rho}(b) (curve 5', dashed line). 

Fig.\,\ref{rho}d shows the derivative of $\rho_\parallel(T)$ with respect to temperature, d$\rho_\parallel$/d$T$. The maxima are observed at $T = 139$\,K and $T = 141$\,K for sample 6 and 5, respectively, which is slightly above the values found for $\rho_\perp$ and probably caused by a different 'background'. 

\subsection{In-plane anisotropy}

\begin{figure}
\includegraphics[width=0.47\textwidth]{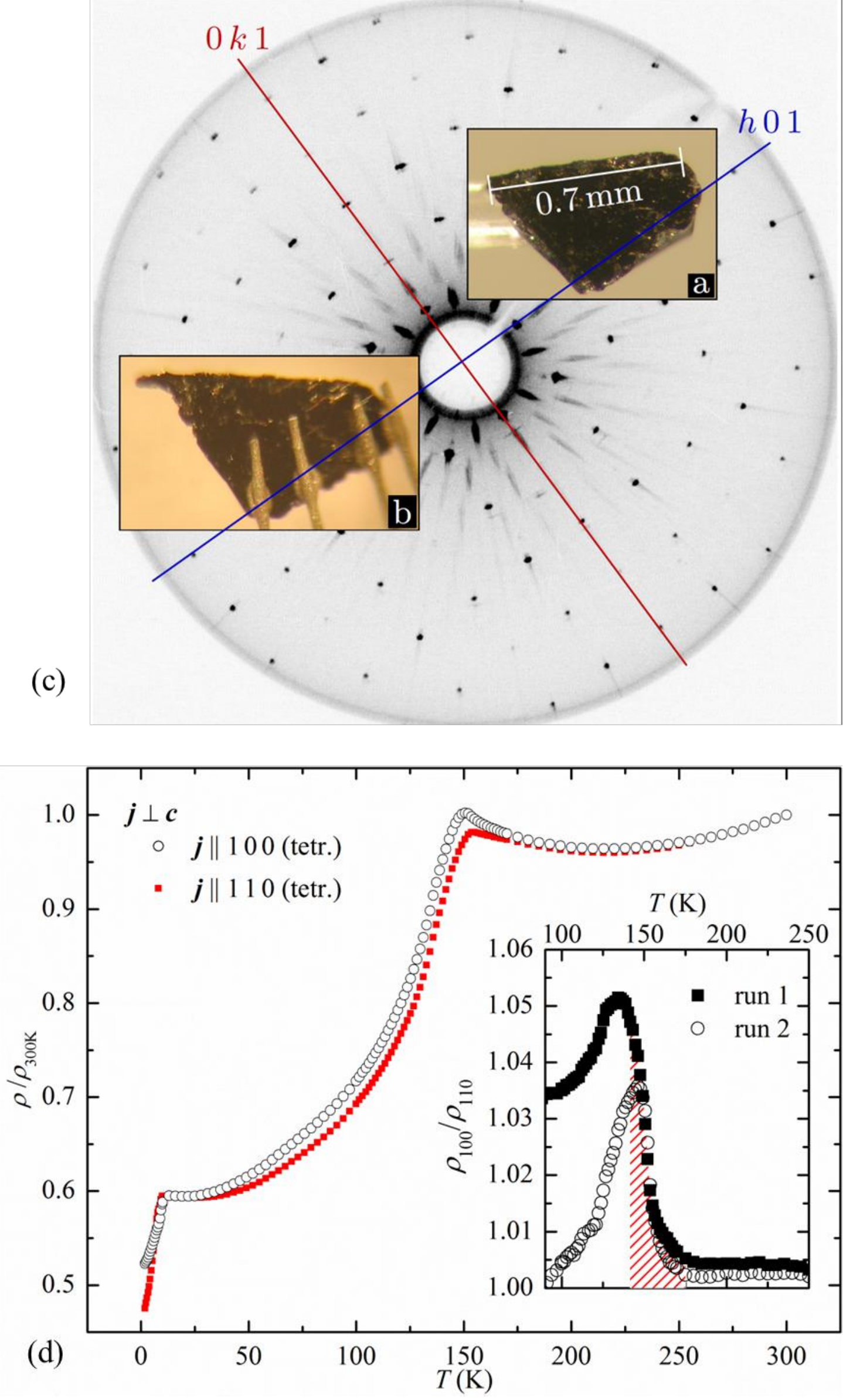}
\caption{(Color online) LaFeAsO single crystal (a) mounted for Buerger precession measurements and (b) with contacts for electrical transport measurements. (c) Precession image of $h\, k\, 1$ reflections (in tetragonal notation). (d) Electrical resistivity for current flow along [100] (open, black circles) and along [110] (filled, red squares). Inset: a significant in-plane anisotropy develops already 30 K above the structural distortion indicating the formation of an electron nematic phase (shaded area). 
}
\label{inplane}
\end{figure}

Rigorously measuring the in-plane anisotropy of the electrical resistivity requires a detwinning of the samples since four types of twin domains form in the orthorhombic, AFM state\,\cite{Tanatar2010}. Therefore, the electrical resistivity of twinned samples is a combination of the $a$- and $b$-axis contributions 
for current flow along the tetragonal $\langle 110 \rangle$ directions or a more complex superposition depending on the in-plane angle for current flow along arbitrary in-plane directions. 
Detwinning of the present LaFeAsO single crystals is technically difficult to achieve due to the relatively small sample size. 
However, it has been shown for BaFe$_2$As$_2$ that even free standing crystals do not develop an equal population of each domain orientation\,\cite{Blomberg2012}. 
Accordingly, a finite in-plane anisotropy of the electrical resistivity can be expected even in twinned samples.

Figure\,\ref{inplane} shows an LaFeAsO single crystal [sample 3 in Fig.\,\ref{rho}(a,c)] mounted for Buerger precession measurements (a) and with contacts for electrical resistivity (b).
The Buerger precession image reveals a clean diffraction pattern consistent with good quality single crystals and easily allows for an in-plane orientation [Fig.\,\ref{inplane}c, depicted is the reciprocal $h\,k\,1$ plane].  
The electrical resistivity was measured with current flow along the [110] direction [$\rho_{110}$, filled, red squares in Fig.\,\ref{inplane}(d)] and along the [100] direction [$\rho_{100}$, open, black circles in Fig.\,\ref{inplane}(d)].
The silver-paint contacts have been dissolved in acetone after the first measurement and new contacts were prepared for the second orientation.
Above $T = 175$\,K the electrical resistivities normalized to RT are almost identical for both orientations. A clear difference develops in the vicinity of \tnfe, \tnull~with a sharper maximum in $\rho_{100}$. At lower temperatures the difference diminishes and the curves cross. 

The inset in Fig.\,\ref{inplane}d shows the temperature-dependence of the in-plane anisotropy defined by the ratio $\rho_{100}$/$\rho_{110}$ for two measurements of $\rho_{100}$ (the second measurement of $\rho_{110}$ revealed an identical resistivity within the experimental
resolution). Even though the two measurements of $\rho_{100}$ are very similar, the calculation of the ratio $\rho_{100}$/$\rho_{110}$ revealed differences. 
The maximum in-plane anisotropy was found at $T = 135$\,K and $T = 147$\,K for run 1 and run 2, respectively, which is in the vicinity of the structural and magnetic phase transitions. 
% Rather, there is good agreement between the position of the maximum in $\rho_{100}$/$\rho_{110}$ and \tnull = 147\,K. 
% The maximum in-plane anisotropy occurs therefore at the orthorhombic distortion rather than at the AFM ordering.
% A small change of slope in $\rho_{100}$/$\rho_{110}$ at $T = 137$\,K indicates an only minor influence of the Fe-AFM ordering on the in-plane anisotropy.

The onset of the in-plane anisotropy takes place at $T = 175$\,K, which is 15\,K above $T = 160$\,K where no indications for an orthorhombic distortion are observable in x-ray diffraction pattern [Fig.\,\ref{diffr}(c)]. Furthermore it is 20\,K above the sharp onset of the peak in the derivative of the electrical resistivity [see Fig.\,\ref{rho}(c) sample 3 and 4] and significantly above the ordering temperature of the (static) structural distortion at \tnull~= 147\,K.
A different $c$-axis contribution in each direction as the origin of the in-plane anisotropy can not account for the observed behavior since there is no corresponding anomaly in $\rho_\parallel(T)$ at $T \sim 175$\,K [Fig.\,\ref{rho}(b)]. Furthermore, $\rho_\parallel$ is increasing under cooling below RT whereas $\rho_\perp$ is decreasing for all measured samples. Therefore, a different $\rho_\parallel$ contribution in $\rho_{100}$ and $\rho_{110}$ of sample 3 would become apparent already between $T = 200$ and 300\,K which is in strong contrast to the almost identical resistivity values observed in this temperature range and the sharp onset of the in-plane anisotropy below $T \sim 175$\,K.

\subsection{Magneto-resistance}

\begin{figure}
\includegraphics[width=0.48\textwidth]{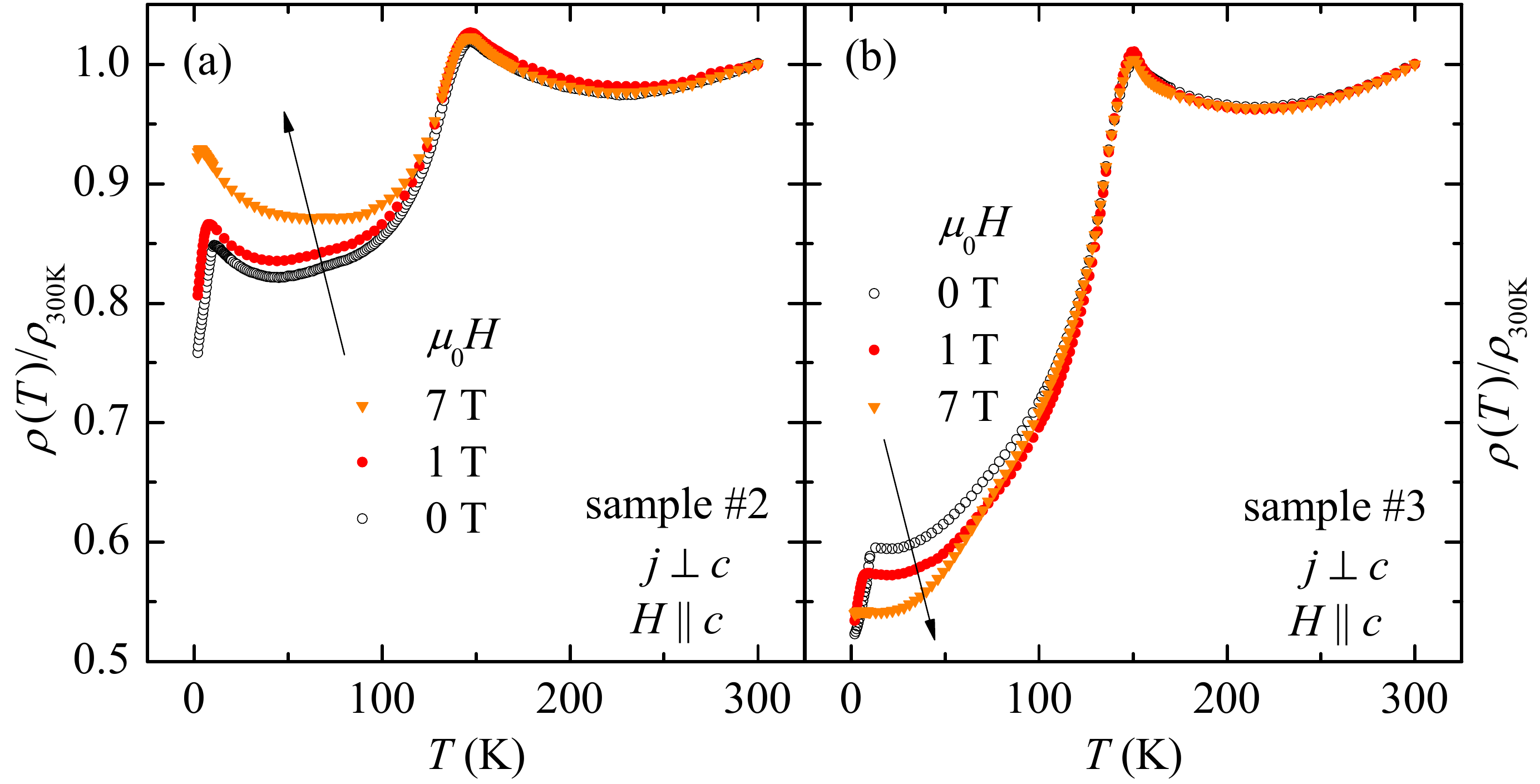}
\caption{(Color online) Magnetic field dependence of the normalized electrical resistivity of LaFeAsO. The magnetoresistance is positive for semi-conducting low-$T$ behavior (a) but can be negative in case of the metallic ground state (b). Note, the magnetoresistance of sample 4 (metallic ground state) was found to be small but positive (not shown). 
}
\label{rho-mag}
\end{figure}

Although the resistivities $\rho_\perp(T)$ of samples 2 and 3 in Fig.\,\ref{rho}(a) behave similar in zero magnetic field, they have opposing response to an applied magnetic field.
Measurements on both samples were performed with the external magnetic field applied along the crystallographic $c$-direction. 
As shown in Fig.\,\ref{rho-mag}a,b the magneto-resistance of sample 2 at low $T$ is positive whereas sample 3 exhibits a negative magneto-resistance. 
$\rho_\perp(T)$ of sample 2 is increasing under cooling below $T = 50$\,K where the minimum shifts to higher $T$ with applied external field. 
Sample 3 shows a metallic resistivity and accordingly a positive magneto-resistance is expected which is in contrast to the observed behavior. Furthermore, the curves measured in $\mu_0H = 1$\,T and $\mu_0H = 7$\,T cross at $T = 75$\,K indicating two competing effects on $\rho(T,H)$. 

In both samples the magneto-resistance increases with decreasing temperature for $T <~$\tnfe,\,\tnull~and is very small for $T >~$\tnull. This indicates that both the positive and the negative magneto-resistance are correlated with the Fe-AFM ordering and the structural distortion.
However, we were not able to identify the origin of the opposing response to a magnetic field and a further discussion of this effect is beyond the scope of this publication.
Note, the magnetoresistance of sample 4 was found to be small but positive (not shown). A metallic ground state is therefore not necessarily connected with a negative magnetoresistance.

The magnetic field dependence of $T^*$ (change of AFM structure of Fe) is almost identical for both samples and manifests a shift of the transition to lower temperatures with increasing field. 

\section{Magnetization}\label{Magnetization}

\begin{figure}
\includegraphics[width=0.47\textwidth]{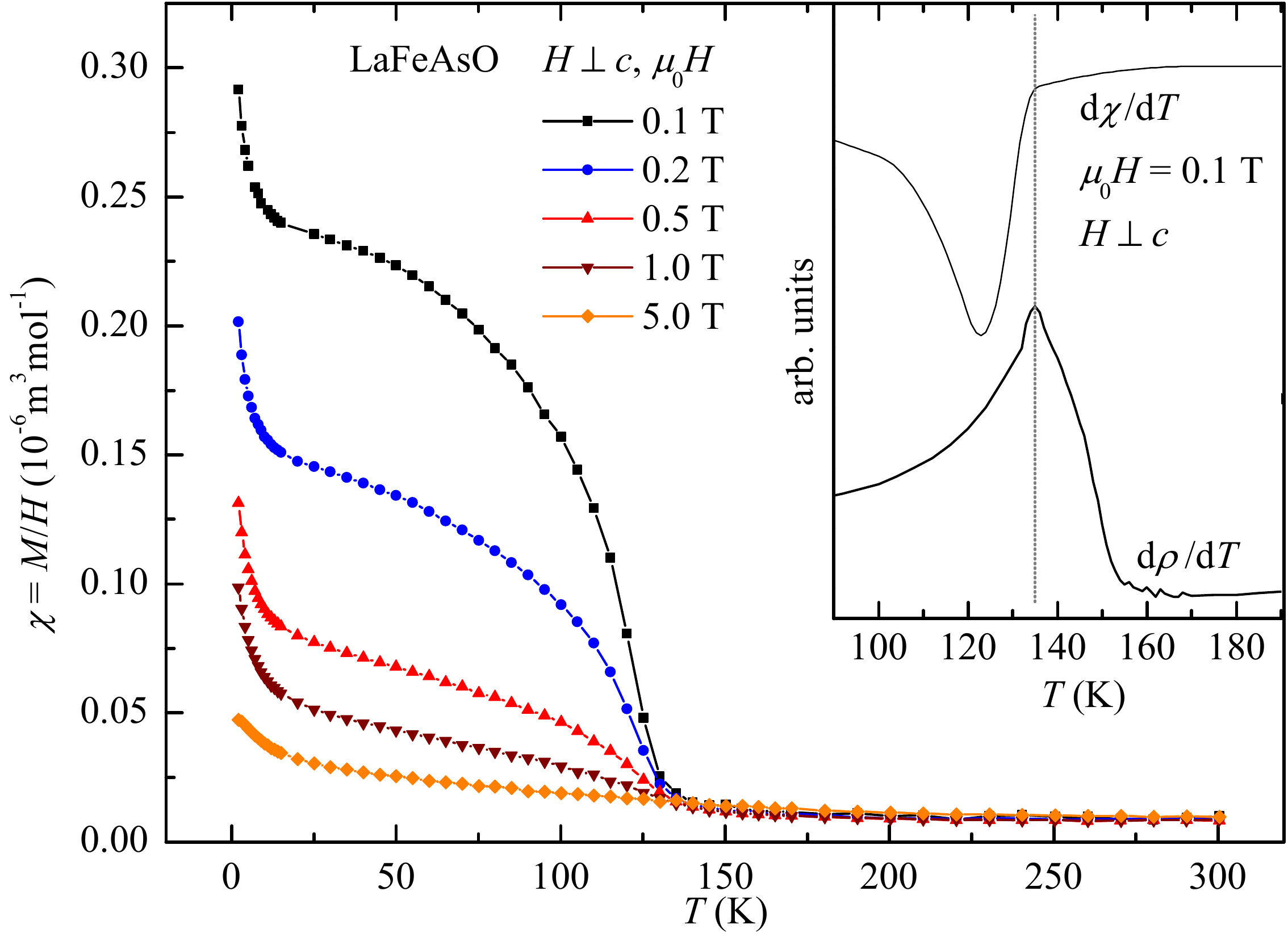}
\caption{(Color online) Magnetic susceptibility $\chi_\perp = M/H$ of LaFeAsO for $\bm H \perp \bm c$. A canting of the Fe-moments causes $\chi_\perp(T)$ to increase below \tnfe. The increase in $\chi_\perp(T)$ below $T = 10$\,K is in accordance with the proposed change of the magnetic structure of Fe from AFM to FM arrangement along the $c$-axis. 
Inset: derivative of the magnetic susceptibility d$\chi_\perp$/d$T$ and of the electrical resistivity d$\rho_\perp$/d$T$. The sharp onset of the anomaly in d$\chi_\perp$/d$T$ at \tnfe~indicates that the canting of the Fe moments develops only in the static AFM state and not in the nematic phase. 
}
\label{mag}
\end{figure}

Figure\,\ref{mag} shows the magnetic susceptibility, $\chi_\perp = M/H$, of LaFeAsO (sample 4) for magnetic fields of $\mu_0H = 0.1 - 5.0$\,T and $H \perp c$. The small field dependence of $\chi_\perp$ for $T > 150$\,K proves the absence of FM foreign phases with an ordering temperature above RT and the values of $\chi_\perp = 9(1)\,\times 10^{-9}$\,m$^3$mol$^{-1}$ at $T = 300$\,K are in good agreement with previous results of $\chi_\perp = 10.3\,\times10^{-9}$\,m$^3$mol$^{-1}$ obtained on single crystals by Yan\,\textit{et al.}\,\cite{Yan2009}. 

The strong increase in $\chi_\perp$ together with the development of a pronounced field dependence below $T \approx 130$\,K manifests what could be interpreted as FM ordering. 
However, a canting of the Fe-moments in the AFM ordered phase results in similar behavior. 
The size of the ordered/canted moment was calculated to be $2.5 \times 10^{-3}\mu_B$/f.u. (f.u. - formula unit LaFeAsO). Similar observations have been made in polycrystalline LaFeAsO$_{0.85}$F$_{0.1}$ that was assumed to contain a fraction of undoped or weakly doped LaFeAsO and manifested an ordered moment of $\sim 1.5 \cdot 10^{-4} \mu_B$/f.u. \,\cite{Fedorchenko2010}.

A comparison of the derivatives d$\chi_\perp$/d$T$ and d$\rho_\perp$/d$T$ reveals a sudden increase of the magnetization during cooling exactly at the position of the maximum in d$\rho_\perp$/d$T$, i.e., only after the formation of static Fe-AFM ordering at \tnfe~= 135\,K but not already in the nematic phase (inset in Fig.\,\ref{mag}).

The strong increase in $\chi_\perp(T)$ below $T = 11$\,K is in good agreement with the proposed change of the magnetic structure of Fe at $T^* = 11$\,K where the FM arrangement along the $c$-axis leads to an increase of the magnetization perpendicular to the $c$-axis (the Fe-moments lie in the basal plane\,\cite{delaCruz2008}). 
In contrast to the suppression of $T^*$ for $H \parallel c$ (inferred from $\rho_\perp$, see Fig.\,\ref{rho-mag}) $T^*$ remains almost constant for $H \perp c$ and $\mu_0H \leq 1$\,T and is somewhat shifted to higher temperatures as well as broadened for $\mu_0H > 1$\,T.
Therefore, the anisotropy of the field dependence of $T^*$ is also in accordance with the proposed magnetic transition since $H \perp c$ favors the FM arrangement along $c$ whereas $H \parallel c$ suppresses the rearrangement.

\section{Discussion}\label{Discussion}

\subsection{Sample dependence of the electric resistivity}\label{DiscussionA}
The origin for the pronounced sample dependent electrical resistivity of LaFeAsO has not been clarified so far. 
Largely varying results were presented for polycrystalline and single crystalline LaFeAsO samples which were synthesized by different crystal growth procedures. Some of them show a more metallic\,\cite{Shekhar2009,Yan2009-arxiv}, others a more semi-conducting behavior\,\cite{Kamihara2008, Mcguire2008, Yan2009-arxiv}.
Our LaFeAsO single crystals were grown under (almost) identical conditions and cover the whole range from semi-conducting to metallic behavior in a continuous fashion [Fig\,\ref{rho}(a)]. Details of the crystal growth procedure such as temperature profiles, quality of the starting materials, or crucibles are therefore not responsible for the observed sample dependencies.
In the following, we discuss three possible scenarios for the origin of the observed behavior.

\subsubsection{c-axis contribution}
A parasitic $c$-axis contribution to the measured in-plane resistivity would explain the sample dependence above \tnfe,\,\tnull~where $\rho_\parallel$ ($\rho_\perp$) increases (decreases) with decreasing temperature. Depending on the ratio of $\rho_\parallel$ and $\rho_\perp$ a local minimum forms between $T \sim 200$ and 250\,K. 
The resistivity $\rho_\perp$ at lower temperatures around \tnfe,\,\tnull~is strongly affected by the in-plane orientation of the current flow as shown for sample 3.
Note that the difference of the resistivities of sample 3 and 4 in Fig.\,\ref{rho}(a) look somewhat similar to the difference of the resistivities for current flow along [1\,0\,0] and [1\,1\,0] of sample 3 in Fig\,\ref{inplane}(b). However, a closer inspection reveals a smooth and continuous increase of $\rho_{\rm sample 3}$/$\rho_{\rm sample 4}$ setting in at $T > 200$\,K in contrast to the sharp onset of the in-plane anisotropy at $T = 175$\,K. 
The strong increase in $\rho_\perp$ at low temperature observed for sample 1 that exceeds the RT value under cooling can not be described by a $c$-axis contribution since $\rho_\parallel$ at low temperatures stays well below the RT values. Furthermore, $\rho_\parallel$ of both samples 5 and 6 is decreasing under cooling between $T = 125$ and 100\,K whereas $\rho_\perp$ of sample 1 is increasing monotonically in this temperature range. 

Therefore, a clear explanation for the overall temperature dependence of the electrical resistivity based on varying $c$-axis contributions or an in-plane anisotropy could not be found. 
There are at least two significant contributions to the electrical resistivity, one metallic and one semi-conducting. 
The anomalies in $\rho(T)$ at \tnfe,\,\tnull~seem to be superimposed to a sample dependent 'background' and are more or less independent of $\rho(T)$ above and below the transition temperatures of \tnfe~= 135\,K and \tnull~= 147\,K.

\subsubsection{Metal-insulator transition}

The increase in $\rho(T)$ under cooling that was found for some samples of LaFeAsO in this as well as in other investigations is somewhat reminiscent of the cuprate parent compounds which are Mott-Hubbard insulators. Indeed, there is further theoretical evidence for an important role of electronic correlations and the possibility for the vicinity of a metal insulator transition in the iron pnictides, see e.g., Ref.\,\onlinecite{Si2008, Abrahams2011}.
The closeness to such an instability would naturally explain the strong sample dependence found for LaFeAsO with some samples being on the metallic side and others on the insulating side. 
On the other hand, there are other possible contributions to the electrical resistivity, e.g., scattering on magnetic impurities, domain boundaries (of magnetic domains formed in the orthorhombic AFM state), or magnetic fluctuations.

\subsubsection{Structural instability}

Another completely different approach to describe the pronounced sample dependence is the assumption of a structural instability. 
With La being the largest rare earth metal, LaFeAsO might be on the stability limit of the ZrCuSiAs structure type. 
On the basis of structural data obtained for a whole series of $R$FeAsO ($R$ = La, Ce, Pd, Nd, Sm, Gd, Tb)\,\cite{Nitsche2010}, the effective radius of lanthanum (corresponding to the La-As distance) is over-proportionally large compared to the rest of the rare earth metals. 
With the height of the lanthanum atom over the oxygen layer being even smaller than expected from the linear trend (Fig. 10 in Ref.\,\onlinecite{Nitsche2010}), a preference of lanthanum towards oxygen and a corresponding weakening of the bonding between the LaO and FeAs layers can be postulated.
This is supported by the significantly higher $c$/$a$-ratio and the highest anisotropy of the crystal growth (basal plane vs. stacking direction) when compared to other $R$FeAsO compounds. 
A resulting structural disorder on a microscopic scale may lead to the sample dependence of the physical properties. 
However, the experimental verification of this hypothesis is difficult since there is no trivalent ion significantly larger than La (but radioactive actinium). 

Altogether, despite the cause of the sample dependence the signatures at \tnfe, \tnull, and $T^*$ are similar for both the metallic and the semi-conducting samples indicating that the Fe-magnetism is not related to the sample dependent electrical transport behavior (Fig.\,\ref{rho}a). 

\subsection{Anisotropy of the electrical resistivity}

The anisotropy of the electrical resistivity between current flow along the crystallographic $c$-axis and the basal plane can be roughly estimated to lie between $\rho_\parallel/\rho_\perp = 20 - 200$ (at RT). 
This is significantly larger than the values of $\rho_\parallel/\rho_\perp < 10$ found in $A$Fe$_2$As$_2$ compounds\,\cite{Tanatar2009b} ($A$ = Ca, Sr, Ba) and reflects the larger structural anisotropy, i.e., the stronger 2-dimensional character of the $R$FeAsO compounds. We define the ratio of the Fe-plane distance to the distance of nearest neighbor Fe atoms in the Fe-plane to compare the different crystal structures. 
This effective $c$/$a$ ratio amounts to $\surd{2} c/a = 3.1$ for LaFeAsO whereas smaller values of $c/(\surd{2}a) = 2.1, 2.2, {\rm and} ~2.3$ are obtained for the $A$Fe$_2$As$_2$ compounds with $A$ = Ca\,\cite{Ni2008b}, $A$ = Sr\,\cite{Krellner2008b}, and $A$ = Ba\,\cite{Rotter2008b}, respectively. 
Despite the large uncertainties of $\rho_\parallel/\rho_\perp$, there is a clear trend seen as an increase of the resistivity anisotropy with increasing effective $c$/$a$ ratio
: $\rho_\parallel/\rho_\perp \sim 2$ for Ca, to $\rho_\parallel/\rho_\perp \sim 4$ for Sr, Ba\,\cite{Tanatar2009b}, to $\rho_\parallel/\rho_\perp > 20$ for LaFeAsO (at RT).
Exfoliation is a possible origin for a high anisotropy, as pointed out by Tanatar\,\textit{et al.}\,\cite{Tanatar2009b}. This is unlikely in the present study because of the small thickness of the LaFeAsO mesas (0.24\,$\mu$m and 0.29\,$\mu$m).   

\subsection{Electronic nematicity}

A detwinning of single crystals is necessary to determine the difference between electrical resistivity along the orthorhombic $a$- and $b$-axis and several measurements have been performed on doped and un-doped $A$Fe$_2$As$_2$ compounds. However, we are not aware of any data obtained on twinned samples with current flow along two different in-plane directions and the influence of a spontaneous, unequal domain distribution on the electrical transport has not been settled so far. 
Furthermore, disentangling the $a$- and $b$- axis contributions is not essential for the discussion of whether an electron nematic phase may exist in the $R$FeAsO compounds or not. 
Our finding of a clear in-plane anisotropy well above \tnfe,\,\tnull~in twinned LaFeAsO single crystals gives strong evidence for the existence of an electron nematic phase (shaded area in the inset of Fig.\,\ref{inplane}.
The sharp onset of the anomaly at $T = 175$\,K does not mean that nematic fluctuations are absent above this temperature. 
In a simplified picture, it rather shows that the nematic fluctuation rate (which decreases with decreasing temperature approaching static ordering) and the spin-fluctuation scattering rate (which increases with decreasing temperature approaching the ordering temperature) are becoming comparable at this temperature, i.e., the relaxation time associated with scattering from spin fluctuations becomes smaller than the typical timescale of nematic fluctuations.
This picture necessitates the assumption that spin-fluctuation scattering is relevant for the electrical transport which holds true for the presence of significant impurity scattering\,\cite{Fernandes2011}.

\section{Summary}

We found a pronounced sample dependence of the electrical transport of LaFeAsO single crystals which is not reflected in their structural properties and can not be clearly ascribed to varying contributions of $\rho_\perp$ and $\rho_\parallel$. 
The Fe-magnetism is essentially unaffected from these variations of the electrical resistivity that range from semi-conducting to metallic behavior at low temperatures.

A pronounced in-plane anisotropy of the electrical resistivity develops under cooling below $T = 175$\,K and shows a maximum at $T \sim 140$\,K. 
The large temperature difference of $\sim 30$\,K between structural distortion and emergence of in-plane anisotropy gives strong evidence for a symmetry breaking of the electronic system when compared to the underlying crystallographic symmetry, i.e., an electron nematic phase.

Magnetic susceptibility and electrical resistivity give evidence for a change of the magnetic structure of Fe from an AFM to a FM arrangement along the crystallographic $c$-axis at $T^* = 11$\,K. This adds further evidence for a frustrated character of the magnetic exchange in $R$FeAsO compounds along the $c$-axis.

\section{Acknowledgments}

C. Steiner, S. K. Kim, M. A. Tanatar, and R. M. Fernandes are acknowledged for fruitful discussions. P. C. Canfield was happy to read and comment on the manuscript.  
The authors thank P. Scheppan and U. Burkhardt for chemical analysis of the samples and T. Meusel for his help in performing the Buerger precession measurements. Part of A.J.'s work was supported by the U.S. Department of Energy, Office of Basic Energy Science, Division of Materials Sciences and Engineering with research being performed at the Ames Laboratory. Ames Laboratory is operated for the U.S. Department of Energy by Iowa State University under Contract No. DE-AC02-07CH11358. 

% \bibliography{zitate}


\end{document}